\documentclass[twocolumn,amssymb,fleqn,showpacs]{revtex4} 
\usepackage{epsfig,amssymb,amsmath,graphicx,subfigure,hyperref  }  
\usepackage{color}

\newcommand{\be}{\begin{eqnarray}}   
\newcommand{\ee}{\end{eqnarray}}

\begin{document}

\title{The shrinking instability of toroidal liquid droplets in the Stokes flow regime}
\author{Zhenwei Yao and Mark J. Bowick}
\affiliation{Department of Physics, Syracuse University, Syracuse,
New York 13244-1130, USA}

\begin{abstract}
We analyze the stability and dynamics of toroidal liquid droplets.
In addition to the Rayleigh instabilities akin to those of a
cylindrical droplet there is a shrinking instability that is
unique to the topology of the torus and dominates in the limit
that the aspect ratio is near one (fat tori). We first find an
analytic expression for the pressure distribution inside the
droplet. We then determine the velocity field in the bulk fluid,
in the Stokes flow regime, by solving the biharmonic equation for
the stream function. The flow pattern in the external fluid is
analyzed qualitatively by exploiting symmetries. This elucidates
the detailed nature of the shrinking mode and the swelling of the
cross-section following from incompressibility. Finally the
shrinking rate of fat toroidal droplets is derived by energy
conservation.

\end{abstract}
 \maketitle

Liquid droplets of various shapes are ubiquitous in nature pure
and applied. They are found in rain, clouds, paint, lubricants,
inks, dyes and oil \cite{deGennes,Craster} and are being
increasingly exploited in microfluidics \cite{microfluidicRMP}.
The instabilities of liquid droplets have attracted attention
since the beginning of the 19th century~\cite{Eggers,ChandraInstability,sciapplication,experimentinstability1,experimentinstability2,experimentinstability3}.
Early work of Plateau showed that a long cylindrical liquid
droplet is unstable to capillary wave deformations of wavelength
exceeding the droplet circumference. Rayleigh subsequently
determined the most unstable capillary mode by solving the
Navier-Stokes equation~\cite{Rayleigh,Rayleigh196}. Purely planar
liquid droplets are, in contrast, stable since capillary waves
always increase the droplet surface area and hence the free
energy~\cite{Safran}. Droplet instabilities thus probe the
combined influence of the three-dimensional geometry of the
droplets and their surface tension~\cite{Eggers}.

In this paper we study the instability of liquid droplets in the
form of three-dimensional axially symmetric solid tori, inspired
by recent experiments in which bulk liquid tori are created by
extruding water or glycerin through a metallic needle into a
rotating bath of viscous silicone oil~\cite{Pairam}. Thin toroidal
droplets exhibit Rayleigh instabilities analogous to those of the
cylinder~\cite{Rayleigh,Tomotika,McGraw}, with the additional
requirement that the most unstable mode has wavelength $\lambda_c$
commensurate with the outer circumference of the torus. When the
outer circumference is an integer (n) times $\lambda_c$, the
toroidal droplet eventually fissions into n solid spherical
droplets (three-dimensional balls). Thus the  change in topology
of the droplet (solid torus breaks up into n balls) is governed by
a Bohr quantization condition with the final number of balls
playing the role of the principal quantum number n. Toroidal
droplets also exhibit a fundamentally different type of
instability in which the torus shrinks to close its interior hole,
eventually becoming a single ball. This instability is a signature
of the topological character of the torus and does not exist for a
cylinder. Although it is present for a torus of any aspect ratio,
it is preempted by the Rayleigh instability unless the torus is
sufficiently fat (see Appendix B).

The outline of this paper is as follows: we first analyze the shrinking
instability of toroidal droplets by minimizing a free energy
controlled by interfacial surface tension. We then derive the pressure
distribution driving bulk flow of the fluid. The shrinking mode is
then examined in more depth via the Stokes equation, which is the
large Ohnesorge number ($Oh \equiv \eta/ \sqrt{\rho L \sigma}>>1$)
limit of the Navier-Stokes equation.  The biharmonic equation for the stream function determines
the velocity field inside the toroidal liquid droplet. The shrinking
of the droplet and simultaneous swelling of the cross-section from
volume conservation are clearly revealed in the flow.  Finally, we
calculate the shrinking speed by balancing the rate of free energy
gain with the viscous dissipation rate.

\begin{figure}[phtb]
\centering
\includegraphics[width=2.4in]{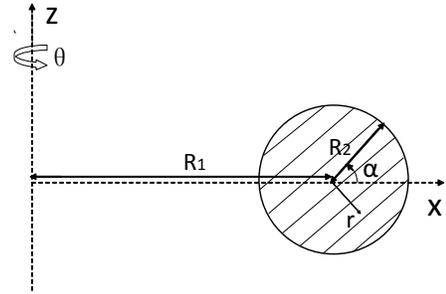}
\caption{Schematic of coordinates for a solid torus generated by
rotating a circular disk of radius $R_2$, centered a distance $R_1$ from the
origin, around the vertical (z) axis.} \label{coord}
\end{figure}

A three-dimensional axially symmetric solid torus is characterized
by coordinates $\{u^1=\alpha,u^2=\theta,u^3=r \}$,  as shown in
Fig.\ref{coord}. Here $\alpha$ is the angle around the tube,
$\theta$ is the azimuthal angle around the z-axis, and $r$ is the
radial coordinate of the tube. The central circle of the solid
torus with radius $R_1$ (at $r=0$) will be called the reference
circle of the solid torus. The outer radius of the tube is denoted
$R_2$. The aspect ratio of the toroidal surface is then
$\phi=R_1/R_2$. The non-zero components of the metric tensor of
the solid torus are $ g_{11}=r^2,\ g_{22}=(R_1 + r \cos\alpha)^2$
and $g_{33}=1$.

The instability of a toroidal liquid droplet to shrinking can be
seen in terms of the free energy $ F=\sigma A$. In the shrinking
process, $R_1$ decreases and $R_2$ increases due to volume
conservation. The change of free energy with radius is \be
\frac{dF}{dR_1}=2\pi^2 \sigma R_2 > 0 .\label{dFt}\ee Thus
toroidal liquid droplets shrink to reduce the free energy.
Although this static analysis reveals the shrinking mode, the free energy
alone does not provide a complete description of the system. In particular,
determining the shrinking rate requires a study of droplet hydrodynamics.

We first analyze the pressure distribution in a toroidal droplet
to understand the driving force for bulk flow. Taking the
divergence of the Navier-Stokes equation for an incompressible
fluid shows that the pressure must be harmonic \be \Delta p(r,
\alpha)=0 \ . \label{laplacianp}\ee The boundary condition is
given by the distribution of Laplace pressure on the interface
between the inner and outer fluids \be p-p_0=\sigma H
,\label{sigmaH}\ee where H is the mean curvature $H=\frac{R_1+2R_2
\cos\alpha}{R_2(R_1+ R_2 \cos\alpha)}$. For simplicity, we first
consider the external pressure $p_0$ as constant. The problem of
solving for the pressure distribution in the bulk fluid is then
reduced to solving Laplace's equation, Eq.($\ref{laplacianp}$),
with the specified boundary condition. The Laplace pressure drop
from the exterior ($\alpha=0$) to the interior ($\alpha=\pi$) of
the torus is given by $P(\alpha=0)-P(\alpha=\pi)=2 \sigma
\frac{1}{R_2} \frac{\phi}{\phi^2-1} $ and is a measure of the
asymmetry of the torus. Since $\phi>1$, the Laplace pressure on
the exterior of the toroid is always bigger than on the interior.
One also sees that the asymmetry is more pronounced for a fat
torus with aspect ratio $\phi$ approaching one. In the limit $\phi
\rightarrow \infty$, a toroid approaches a solid cylinder and the
asymmetry as well as the shrinking mode disappear. Note that for
the opposite case of constant pressure in the inner fluid, the
pressure in the outer fluid will fall from the interior to the
exterior of the torus. The outer fluid will therefore flow outward
and the inner fluid will correspondingly flow inward, shrinking
the droplet. Shrinking is thus a universal feature of one toroidal
fluid inside another.
\begin{figure}[phtb]
\centering
\includegraphics[width=2.5in]{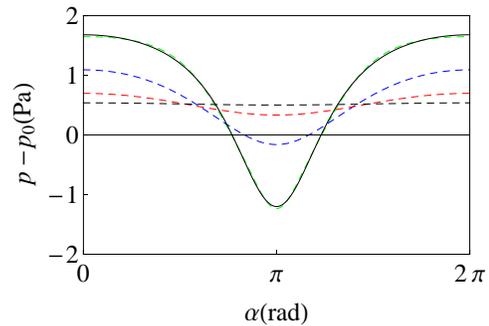}
\caption{The pressure $p(r,\alpha)$ vs. angle $\alpha$ at
different radial distances away from the reference circle for
$R_1=5\ \textrm{mm}$ and aspect ratio $\phi=R_1/R_2=1.5$. The
solid black curve is the pressure distribution on the boundary
$p=\sigma H$. Green curve: $r=R_1/1.5$ (boundary). Blue curve:
$r=R_1/3$. Red curve: $r=R_1/10$. Dashed Black curve: $r=R_1/100$.
The green curve fits the exact pressure on the boundary very
well.} \label{palpha}\end{figure}

Laplace's equation for the pressure Eq.($\ref{laplacianp}$) separates in the coordinates
$\{ \rho, \varphi, \theta \}$\cite{torusgeometry1,EMbooktorus} defined by
\begin{displaymath} \vec{x}(\rho, \varphi, \theta) = \left(
\begin{array}{c}
\frac{a \sinh\rho \cos\theta}{\cosh\rho-\cos\varphi} \\
\frac{a \sinh\rho \sin\theta}{\cosh\rho-\cos\varphi}    \\
\frac{a \sin\varphi}{\cosh\rho-\cos\varphi} \\
\end{array} \right),
\end{displaymath} where $a=r \sinh\rho$.

Exploiting azimuthal symmetry and imposing the requirement that
the pressure be finite as one approaches the reference circle, the
physically acceptable solution takes the form ~\cite{EMbooktorus}
\be p=\sqrt{\cosh\rho-\cos\varphi} \sum_{p\in \mathbb{Z}} \alpha_p
\textrm{Re}[Q_{p-1/2}(\cosh\rho)]\cos (p\varphi)\label{paccept}\ee
where $Re[Q_{p-1/2}(x)]$ is the real part of the associated
Legendre function of the second kind. The coefficients $\alpha_p$
can be determined by imposing the boundary condition in
Eq.($\ref{sigmaH}$). The pressure distribution inside the droplet
is plotted in Fig.$\ref{palpha}$ for aspect ratio $\phi=1.5$,
surface tension $\sigma=4\times 10^{-3}\ \textrm{N}/\textrm{m}$
and $R_1=5\ \textrm{mm}$. The pressure clearly drops from the
exterior ($\alpha=0,2\pi$) to the interior ($\alpha=\pi$). This
pressure gradient drives the fluid towards the center of the
toroid. As $r$ decreases the pressure distribution becomes more
isotropic ($\alpha$-independent). Note that only the $p=0$ mode in
Eq.($\ref{paccept}$) contributes to the pressure near the
reference circle. This enables us to study the behavior of the
fluid near the reference circle analytically. By inserting the
zero mode in Eq.($\ref{paccept}$) into the Stokes equation one can
determine the velocity field near the reference circle in $\{ x,z
\}$ coordinates: $v_x=v_{\alpha} \sin\alpha-v_{r}\cos\alpha =c$,
and $v_z=v_{\alpha} \cos\alpha+v_{r}\sin\alpha=0$. The velocity
field near the reference circle is uniform towards the center of
the toroid.

We now turn to the velocity distribution in a viscous toroidal liquid
droplet. In this regime of large Ohnesorge number the
Navier-Stokes equation reduces to the Stokes equation \be \Delta
\vec{v}=\frac{1}{\eta} \nabla p \label{StksEq},\ee where $\eta$ is
the fluid viscosity. Here viscous dissipation dominates over kinetic energy damping: \be \frac{\dot{E}_{kin}} {\dot{E}_{visc}
}=\frac{d[\frac{1}{2}\int \rho v^2 dV]/dt}{\int dV \sigma'_{ij}
\nabla^i v^j} \propto \frac{\rho L \sigma}{\eta^2}<<1,\ee where $
\sigma'_{ij}=\eta (\nabla_i v_j+\nabla_j v_i)$ is the viscous
stress tensor. Since the characteristic speed of the fluid is much
smaller than the speed of sound, we can treat the fluid as
incompressible ($\nabla \cdot \vec{v}=0$)\cite{Landau}. For
incompressible fluids one can write the velocity field as the curl
of a vector potential $\vec{\psi}$ ($\vec{v}=\nabla \times
\vec{\psi}$) leading directly to \be \Delta^2 \vec{\psi}=0
\label{biharm} .\ee

The complete velocity field can be obtained by solving the
biharmonic vectorial equation Eq.($\ref{biharm}$) which reduces to
a simplified scalar differential equation in the $\{ \rho,
\varphi, \theta \}$ coordinates\cite{E4paperbiharmonic} \be
E^2(E^2\psi)=0 ,\label{E4}\ee where $\psi$, the stream function,
is the only non-zero component $\psi_{\theta}$ of the vector
potential $\vec{\psi}$ and the second-order partial differential
operator $E$ is given by $ E^2=w h^2[
\partial_{\rho}(\frac{1}{w}\partial_{\rho})+
\partial_{\varphi}(\frac{1}{w}\partial_{\varphi})]$, with $w=a\
\sinh (\rho )/[\cosh (\rho )-\cos (\varphi )]$ and $h=[\cosh (\rho
)-\cos (\varphi )]/a $. Imposing the physical requirements that
approaching the reference circle $v_x$ tends to a finite value and
$v_z \rightarrow 0$ (reflection symmetry) yields the complete
solution \be \psi=\frac{a
\sinh\rho}{(\cosh\rho-\cos\varphi)^{3/2}}
\sum_{\nu=-\infty}^{+\infty} c_{\nu} \sin(\nu \varphi)
Q_{\nu-3/2}^1(\cosh\rho) \label{ansatzpsi}\ee Note that $v_x
\rightarrow -\frac{\pi c_1}{2 \sqrt{2}}$ and $v_z \rightarrow 0$
as $\rho \rightarrow \infty$ ($r \rightarrow 0$). Thus only the
$\nu=1$ mode contributes to the flow near the reference circle.
The coefficients $c_{\nu}$ in Eq.($\ref{ansatzpsi}$) can be
determined by matching to the velocity field on the interface.
Assuming that high viscosity fixes the fluid particles on the
interface to move with the interface as it shrinks, the boundary
conditions are found to be $v_{x0}=V(1-\frac{x_0}{2\ R_{1}})$ and
$v_{z0}=-\frac{V\ z_0}{2\ R_{1}}$, where $x_0$ and $z_0$ denote
spatial points on the boundary and $V\equiv dR_{1}/dt$. The point
$(x=0,z=0)$ is the center of the cross section. The $\nu=1$ mode
is sufficient to fit this boundary condition.

\begin{figure}
\includegraphics[width=3in,bb=18 141 711 470]{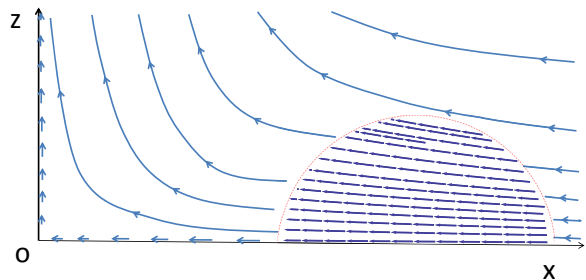}
\caption{The velocity field inside and outside a cross-section of a toroidal
liquid droplet. The dashed semi-circle is the interface of two distinct fluids.
The velocity field inside the droplet is calculated by solving the biharmonic equation.
The external flow pattern is schematically plotted by imposing boundary conditions and exploiting symmetry.
Parameters are mode number $\nu=1$, $R_{1}=5\ {\rm mm}$ and $R_2=2\ {\rm mm}$.}
\label{va1}\end{figure}

\begin{figure}
\includegraphics[width=2.0in]{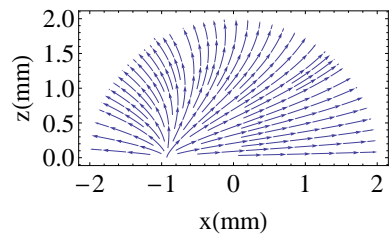}
\caption{The velocity field inside the toroidal droplet of
Fig.$\ref{va1}$ in a comoving reference frame shrinking with the droplet. The swelling of the cross section is readily inferred. }
\label{vab}\end{figure}

The velocity field inside the droplet in the laboratory frame is
plotted in Fig.$\ref{va1}$. The shrinking of the droplet is
clearly indicated by the inward directed flow inside the droplet.
One also sees that outer fluid within the toroidal hole is
squeezed out. Further insight is provided by plotting the velocity
field (see Fig.$\ref{vab}$) inside the droplet in a reference
frame comoving with the shrinking of the droplet. The swelling of
the cross section resulting from volume conservation is clearly
visible.

\begin{figure}[tbp]
\centering\subfigure[]{
\includegraphics[width=2.0in]{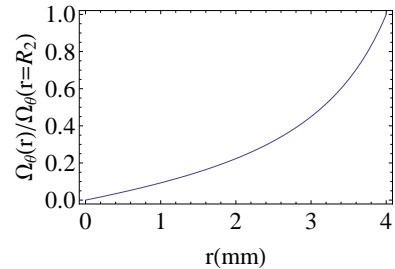}}
\subfigure[]{
\includegraphics[width=2.2in]{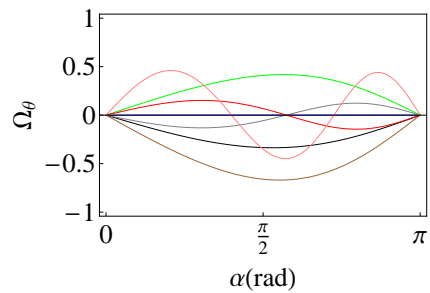}}
\caption{(a) The vorticity
$\Omega_{\theta}(r)/\Omega_{\theta}(r=R_2)$ versus $r$.
Parameters: $\nu=1$, $c_1=1$, $\alpha=1$ rad, $R_1=5$ {\rm mm} and
$R_2=4 \textrm{mm}$ . The vorticity field falls to zero at the
reference circle. (b) $\Omega_{\theta}$ versus angle $\alpha$ for
modes $\nu$ in the range $(-3,3)$. The vorticity $\Omega_{\theta}$
is rescaled to show different modes in the same figure.
Parameters: $\phi=5,a=1$. Blue curve: $\nu=0$ (vorticity
vanishes). Black curve: $\nu=1$. Brown curve: $\nu=2$. Gray curve:
$\nu=3$. Green curve: $\nu=-1$. Red curve: $\nu=-2$. Pink curve:
$\nu=-3$. Note that the number of peaks or valleys is determined
by the mode number $\nu$. } \label{vorticityr}
\end{figure}

The vorticity field, $\vec{\Omega}\equiv\nabla \times
\vec{v}=-\Delta \vec{\psi}$ is plotted in Fig.$\ref{vorticityr}$
which shows its only non-zero component $\Omega_{\theta}$ as a
function of $r$ and $\alpha$ respectively.
Fig.$\ref{vorticityr}$(a) shows that the vorticity field is only
significant near the boundary {--} it decays rapidly as one
approaches the reference circle. Fig.$\ref{vorticityr}$(b) shows
$\Omega_{\theta}$ versus $\alpha$ for $\nu \in [-3,3]$. The
vorticity field vanishes at $\alpha=0$ and $\pi$ due to its odd
parity. The sign of $\Omega_{\theta}$ reflects the chiral property
of vortices. The number of peaks and valleys on the $z>0$ plane
(i.e., $\alpha \in [0,\pi]$) is
\begin{displaymath}
n= \left\{ \begin{array}{ll}
-\nu, & \nu<0\\
\nu-1, & \nu>1\\
1, & \nu=1,
\end{array} \right.
\end{displaymath}
and is therefore completely determined by the mode $\nu$.

In the process of shrinking the free energy gained is dissipated
in viscous damping. By equating the rate of change of the free
energy, Eq.($\ref{dFt}$), to the viscous dissipation rate we can
obtain the shrinking speed. We focus here on the experimentally
explored case of a low viscosity ($\eta_i$) inner fluid immersed
in a viscous ($\eta_o$) outer bath~\cite{Pairam}. In this case the
dissipation occurs almost entirely in the outer fluid. Applying
Stokes' equation for an incompressible fluid the dissipation rate
can be separated into two parts \be \dot{E}_{vis}=-\int dV
\sigma'_{ij}
\partial_i v_j =-\int df_i \sigma'_{ij}v_j \\\nonumber +\int dV
v_j\partial_i \sigma'_{ij} \ . \label{Evs}\ee Here $df_i$ is the
$i$ component of the area element of the interface. The first term
is the heat flux on the fluid boundary and the second term is the
dissipation rate inside the bulk fluid. The second term can be
related to the vorticity: $\int v_j\partial_i \sigma'_{ij} dV=\eta
\int v_j\Delta  v_j dV=-\eta \int \vec{v}\cdot (\nabla\times
\vec{\Omega})dV$. Since the Reynolds number of the external fluid
is very small in the experimental setup ($Re \approx
10^{-4}$~\cite{Pairam}), we may take the external flow to be as
irrotational (vanishing vorticity) by recalling the experiment of
flow through a cylindrical solid: an irrotational-rotational flow
transition occurs at $Re\sim 1$, below which the flow is
irrotational~\cite{Feynman}. In the shrinking process, the
toroidal droplet moves in the external fluid. This is equivalent
to flow through the toroidal droplet. Since the viscosity of the
internal fluid is very small in this case, the internal
dissipation can be neglected. Thus we need to calculate only the
surface integral in Eq.(10) to obtain the dissipation rate. We
need the viscous stress on the boundary to evaluate the surface
integral. Rotational symmetry and the limiting condition
$\eta_i/\eta_o \ll 1$ imply both $\sigma'_{r\theta}$ and
$\sigma'_{r\alpha}$ vanish at the interface. To determine
$\sigma'_{rr}=2 \eta_o
\partial_{r} v_{r}$ we need the gradient of $v_{r}$ at the interface.
Assuming that the fluid particles \textit{near} the boundary move together with the interface
during shrinking, as a result of the viscous external fluid, we have \be
\partial_{r} v_{r}|_{interface}=\frac{6  R_{1}}{R_2^2}
\dot{R_2} \cos\alpha .\label{vgrad} \ee Inserting Eq.(11) along
with the velocity field on the interface into the surface term of
Eq.(10) yields \be \dot{E}_{vis}=-24 \pi^2 \eta_o
((\frac{R_1}{R_2})^2-\frac{1}{2}) R_2\dot{R}_1 \dot{R}_{2} \ .
\label{Evfin}\ee

\begin{figure}
\includegraphics[width=2.2in]{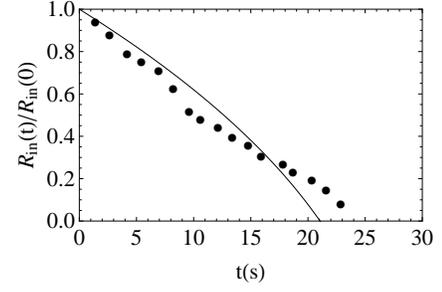}
\caption{The time evolution of the normalized inner droplet
radius $R_{in}(t)/R_{in}(0)$ for initial aspect ratio $1.4$.
The theoretical result is the solid curve and the experimental data points are taken
from Ref.\cite{Pairam}. Parameters: $R_1(0)$= 3
{\rm mm}; $ v_s= \sigma/\eta_o = 133\ \mu{\rm m/s}$.}
\label{sspeed19}\end{figure}

By equating the rate of change of the free energy from
Eq.($\ref{dFt}$) and the dissipation rate Eq.($\ref{Evfin}$), we
have \be \dot{R}_{2}(t)=\frac{v_s}{12}  \frac{1}{\phi^2(t)-1/2}
\label{dR20} \ee and the interior hole of the droplet decreases in
size according to \be \dot{R}_{in}(t)=-\frac{v_s}{12}  \frac{2\
\phi(t)+1}{\phi^2(t)-1/2} ,\label{dRin}\ee where
$\phi(t)=R_1(t)/R_2(t)$ and $v_s=\sigma/\eta_o$. The shrinking
speed is controlled by the aspect ratio of the droplet, in accord
with experimental observations~\cite{Pairam}. In the limit of
infinite aspect ratio (the cylinder) the shrinking speed vanishes,
as required. The constant $1/2$ in the denominator of
Eq.(\ref{dR20}) plays an important role in the limit that the
aspect ratio approaches one (fat tori). The plot of
$R_{in}(t)/R_{in}(0)$ versus $t$ is shown in Fig.$\ref{sspeed19}$
for an initial aspect ratio $R_{1}(0)/R_2(0)=1.4$,
$\eta_i/\eta_o=1/30,000$, $\sigma=4\ \textrm{mN/m}$,
$R_{1}(0)\approx\ 3\ \textrm{mm}$ and $\eta_i=10^{-3}\
\textrm{kg}/(\textrm{m} \cdot \textrm{s})$. For these parameters $
v_s= \sigma/\eta_o \approx 133\ \mu \textrm{m}/\textrm{s} $.
Fig.$\ref{sspeed19}$ shows that droplets shrink with roughly
constant speed, as found in~\cite{Pairam}. Our results predict
that it would take about $21\ \textrm{s}$ (aspect ratio $1.4$) and
$50\ \textrm{s}$ (aspect ratio $1.9$) for a toroidal droplet to
shrink to close the inner hole of toroid, in qualitative agreement
with the experimental values of $25\ \textrm{s}$ and $38\
\textrm{s}$ respectively. Thus thinner toroidal droplets shrink
more slowly, consistent with experimental observations
\cite{Pairam}.

Our energy conservation approach to determining the shrinking
speed can also be applied to a 2-dimensional system where it
yields an analytical result. Consider a shrinking hole on a liquid
film. The limiting case of a shrinking toroidal liquid droplet
with $R_{in} \rightarrow 0$ and $\eta_i/\eta_o>>1$ can be modelled
as such a 2-dimensional system, since the dynamics of the hole
becomes independent of the fluid far away from the hole. As the
hole shrinks, a flow will be induced outside the hole on the film.
In the Stokes flow regime, the velocity field can be derived
analytically in the polar coordinates $\{ \rho,\theta \}$ as
$v_{\rho}=\frac{1}{\rho}r(t)\dot{r}(t)$ and $v_{\theta}=0$. By
energy conservation, the shrinking speed of the hole can also be
derived analytically. By equating the rate of change of the line
energy $ \dot{E}_s=\frac{dE_s}{dt}=2\pi \gamma \ \dot{r} $ and the
viscous dissipation rate $ \dot{E}_{vis}=-\int \rho d\rho d\theta
(\sigma'^{\rho\rho} \partial_{\rho}v_{\rho})=-2\pi\eta\ \dot{r}^2
$, we have $\dot{r}=\frac{\gamma}{\eta}$, where $\eta$ is the
viscosity of fluid and $\gamma$ is the line tension.

We expect that the formalism employed here will have a variety of applications
to the dynamics of fluid interfaces. It may also be extended to liquid crystalline droplets where the
interplay of liquid crystalline order and the shape of the droplet should be very rich.

\section*{Acknowledgements}

We thank Xiangjun Xing for extensive discussions and Alberto
Fern{\'a}ndez-Nieves for introducing us to his beautiful
experiments on toroidal droplets. This work was supported by the
National Science Foundation grant DMR-0808812 and by funds from
Syracuse University.

\section*{Appendix}

\numberwithin{equation}{subsection}
\subsection{The shape of cross section}


So far we have assumed that droplets remain perfectly circular in cross-section as they shrink.
Here we show that this assumption is well justified.

\begin{figure}[htb]
\centering \subfigure[]{
\includegraphics[width=1.1in]{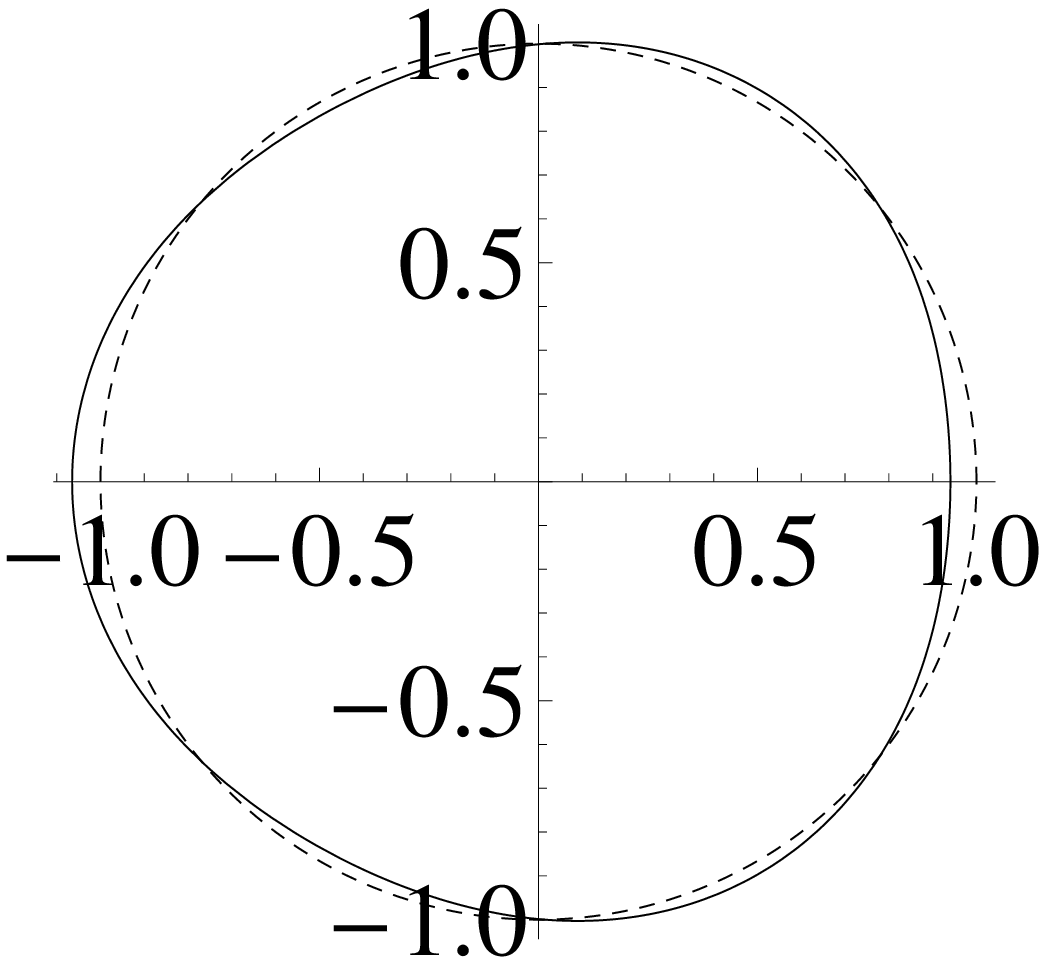}}
\hspace{0.3in} 
\subfigure[]{
\includegraphics[width=1.1in]{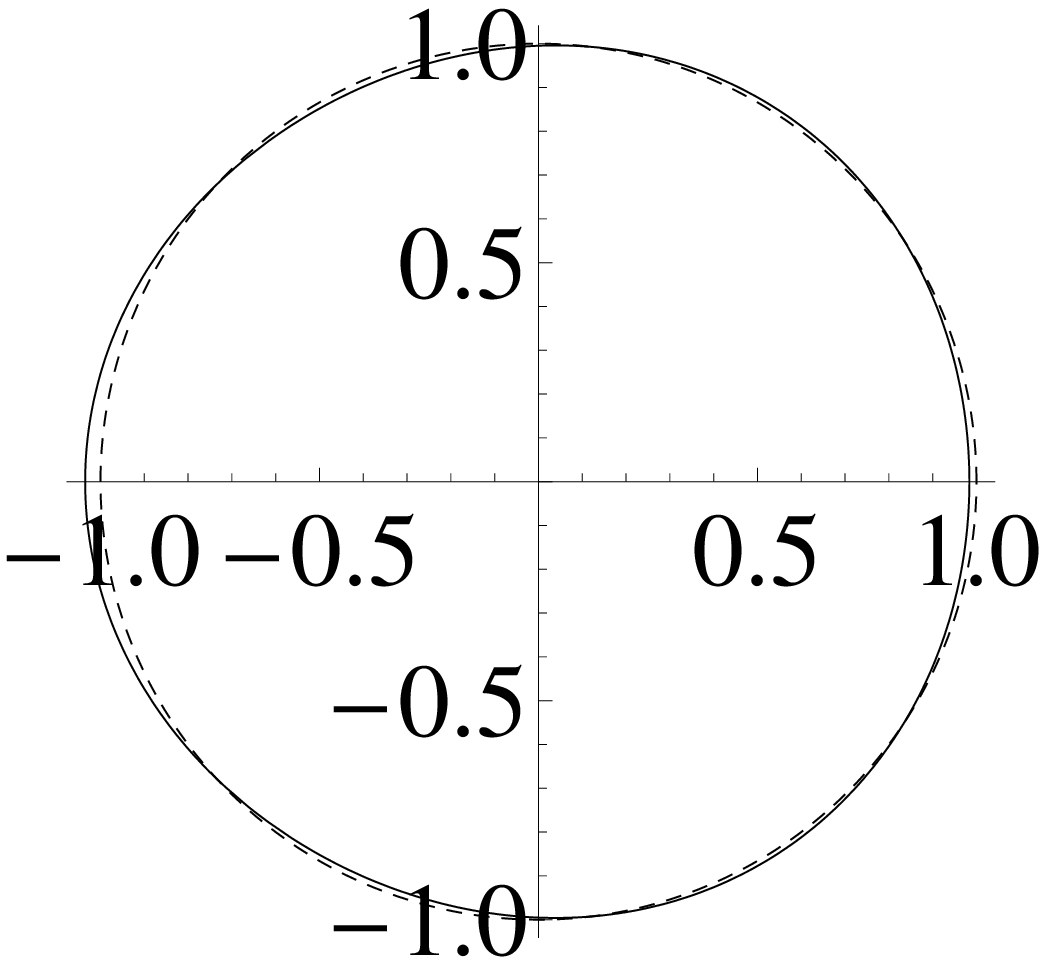}}
\caption{Minimum energy cross-sectional shapes within  a two
parameter family of possible shapes (see Eq.($\ref{ansatcs}$)).
The dashed curves are the unperturbed circular shapes: (a) aspect
ratio 10 (thin torus) and (b) aspect ratio 2 (fat torus).
Parameters: $R_2=1$.} \label{shape_cs}\end{figure}

The shape of a toroidal liquid droplet is characterized by the
radii $ R_1$ and $R_2$ which may in general vary with $\alpha$ and
$\theta$. Retaining azimuthal symmetry we consider the following
ansatz for $R_2$ at a fixed time: \be R_2(\alpha)=a +
c_2P_2(\cos\alpha) + c_3 P_3(\cos\alpha).   \label{ansatcs} \ee
The second term describes an ellipse which is symmetric about z
axis, while the third term describes a shape with three round
corners, which is asymmetric about the z-axis (we are ignoring the
shrinking mode here, described by a $P_1(\cos\alpha)$ term). The
shape of the droplet is specified by points in the $\{c_2,
c_3|c_2, c_3 \in [-b,b]  \}$ space.

We numerically search for the ground state in the $\{c_2, c_3|c_2,
c_3 \in [-b,b]  \}$ space for which \be L=A-A_0+\lambda |V-V_0|
\ee is minimized, where $V_0$ and $A_0$ are the volume and surface area of
the unperturbed droplet. $\lambda$ is set to be large to impose volume conservation. We take
$a=1$ and $b=0.2$.

For tori with typical aspect ratios $R_1/R_2=10$ and $2$, we find
the ground states in the $\{c_2, c_3|c_2, c_3 \in [-b,b]  \}$
space shown in Fig.$\ref{shape_cs}$. The cross-sections are very close to circular.
In the experimental work of \cite{Pairam} this is also true.

\subsection{Rayleigh instability vs shrinking mode}

It is observed experimentally\cite{Pairam} that the Rayleigh
instability disappears for sufficiently fat solid tori
($R_1(t=0)/R_2(t=0) \lesssim 2$) whereas the shrinking mode is
present for all aspect ratios. Here we derive a lower bound on the
aspect ratio for the emergence of the Rayleigh instability.

Two conditions must be satisfied for the Rayleigh instability: (1)
modes with wavelength $\lambda> \lambda_c$, where $\lambda_c$ is
the minimum wavelength of the Rayleigh instability mode and  (2):
\be u_k(t=t_1)\gtrsim R_2(t=0),\label{criterion}\ee where $u_k$ is
the perturbation amplitude and $t_1$ is the lifetime of the
shrinking droplet ($R_{in}(t=t_1)=0$).  It is well
known~\cite{Rayleigh,Rayleigh196} that $u_k$ grows exponentially:
\be u_k(t)=u_k(0)e^{v_s t/R_2(0)} ,\label{uk}\ee  where
$v_s=\sigma/\eta$ is the characteristic speed and $R_2(0)$ the
characteristic length scale of the system. We assume that $u_k$
grows exponentially all the way until breakup of the droplet. On
the other hand we have shown that $R_{in}(t)$ decreases almost
linearly in time. So formally, we have \be
R_{in}(t)=R_{in}(t=0)-v_c t ,\label{RinRayleigh}\ee from which we
have $t_1=\frac{R_{in}(t=0)}{v_c}$. By inserting
Eq.($\ref{uk}$,$\ref{RinRayleigh}$) into Eq.($\ref{criterion}$),
we obtain \be \frac{R_1(0)}{R_2(0)} \gtrsim
1+c\ln(\frac{R_2(0)}{u_k(0)}) ,\label{criterion2}\ee where
$c=v_c/v_s$ is an aspect ratio factor of order one that tends to
$1/2$ for aspect ratio one by Eq.(\ref{dRin}). Thus the Rayleigh
instability is dominant for sufficiently thin tori.

It can be checked that for aspect ratios satisfying
Eq.($\ref{criterion2}$) even the perimeter of the interior of the
torus ($2\pi (R_1(0)-R_2(0))$) can accommodate the Rayleigh
instability mode, i.e., \be \frac{2\pi
(R_1(0)-R_2(0))}{R_2(0)}\gtrsim 2\pi \ln(\frac{R_2(0)}{u_k(0)}) >
2\pi .\ee

\end{document}